\documentclass[10pt,conference]{IEEEtran}
\usepackage[latin1]{inputenc}
\usepackage{float}
\usepackage{tabstackengine}
\usepackage{url}
\usepackage[numbers,sort&compress]{natbib}
\usepackage[linesnumbered]{algorithm2e}
\usepackage{tikz}
\usetikzlibrary{shapes, arrows}

\usepackage{listings}
\usepackage{xcolor}
\definecolor{maroon}{cmyk}{0, 0.87, 0.68, 0.32}
\definecolor{halfgray}{gray}{0.55}
\definecolor{ipython_frame}{RGB}{207, 207, 207}
\definecolor{ipython_bg}{RGB}{247, 247, 247}
\definecolor{ipython_red}{RGB}{186, 33, 33}
\definecolor{ipython_green}{RGB}{0, 128, 0}
\definecolor{ipython_cyan}{RGB}{64, 128, 128}
\definecolor{ipython_purple}{RGB}{170, 34, 255}

\lstdefinelanguage{iPython}{
    morekeywords={access,and,break,class,continue,def,del,elif,else,except,exec,finally,for,from,global,if,import,in,is,lambda,not,or,pass,print,raise,return,try,while},%
    morekeywords=[2]{abs,all,any,basestring,bin,bool,bytearray,callable,chr,classmethod,cmp,compile,complex,delattr,dict,dir,divmod,enumerate,eval,execfile,file,filter,float,format,frozenset,getattr,globals,hasattr,hash,help,hex,id,input,int,isinstance,issubclass,iter,len,list,locals,long,map,max,memoryview,min,next,object,oct,open,ord,pow,property,range,raw_input,reduce,reload,repr,reversed,round,set,setattr,slice,sorted,staticmethod,str,sum,super,tuple,type,unichr,unicode,vars,xrange,zip,apply,buffer,coerce,intern,True,False},%
    sensitive=true,%
    morecomment=[l]\#,%
    morestring=[b]',%
    morestring=[b]",%
    morestring=[s]{'''}{'''},
    morestring=[s]{"""}{"""},
    morestring=[s]{r'}{'},
    morestring=[s]{r"}{"},%
    morestring=[s]{r'''}{'''},%
    morestring=[s]{r"""}{"""},%
    morestring=[s]{u'}{'},
    morestring=[s]{u"}{"},%
    morestring=[s]{u'''}{'''},%
    morestring=[s]{u"""}{"""},%
    literate=
    {á}{{\'a}}1 {é}{{\'e}}1 {í}{{\'i}}1 {ó}{{\'o}}1 {ú}{{\'u}}1
    {Á}{{\'A}}1 {É}{{\'E}}1 {Í}{{\'I}}1 {Ó}{{\'O}}1 {Ú}{{\'U}}1
    {à}{{\`a}}1 {è}{{\`e}}1 {ì}{{\`i}}1 {ò}{{\`o}}1 {ù}{{\`u}}1
    {À}{{\`A}}1 {È}{{\'E}}1 {Ì}{{\`I}}1 {Ò}{{\`O}}1 {Ù}{{\`U}}1
    {ä}{{\"a}}1 {ë}{{\"e}}1 {ï}{{\"i}}1 {ö}{{\"o}}1 {ü}{{\"u}}1
    {Ä}{{\"A}}1 {Ë}{{\"E}}1 {Ï}{{\"I}}1 {Ö}{{\"O}}1 {Ü}{{\"U}}1
    {â}{{\^a}}1 {ê}{{\^e}}1 {î}{{\^i}}1 {ô}{{\^o}}1 {û}{{\^u}}1
    {Â}{{\^A}}1 {Ê}{{\^E}}1 {Î}{{\^I}}1 {Ô}{{\^O}}1 {Û}{{\^U}}1
    {œ}{{\oe}}1 {Œ}{{\OE}}1 {æ}{{\ae}}1 {Æ}{{\AE}}1 {ß}{{\ss}}1
    {ç}{{\c c}}1 {Ç}{{\c C}}1 {ø}{{\o}}1 {å}{{\r a}}1 {Å}{{\r A}}1
    {€}{{\EUR}}1 {£}{{\pounds}}1
    {^}{{{\color{ipython_purple}\^{}}}}1
    {=}{{{\color{ipython_purple}=}}}1
    {+}{{{\color{ipython_purple}+}}}1
    {*}{{{\color{ipython_purple}$^\ast$}}}1
    {/}{{{\color{ipython_purple}/}}}1
    {+=}{{{+=}}}1
    {-=}{{{-=}}}1
    {*=}{{{$^\ast$=}}}1
    {/=}{{{/=}}}1,
    literate=
    *{-}{{{\color{ipython_purple}-}}}1
     {?}{{{\color{ipython_purple}?}}}1,
    identifierstyle=\color{black}\ttfamily,
    commentstyle=\color{ipython_cyan}\ttfamily,
    stringstyle=\color{ipython_red}\ttfamily,
    keepspaces=true,
    showspaces=false,
    showstringspaces=false,
    rulecolor=\color{ipython_frame},
    frame=single,
    framexleftmargin=-2mm,
    framexrightmargin=-2mm,
    numbers=none,
    numberstyle=\tiny\color{halfgray},
    backgroundcolor=\color{ipython_bg},
    basicstyle=\scriptsize,
    keywordstyle=\color{ipython_green}\ttfamily,
}

\title{QuNetSim: A Software Framework for Quantum Networks}

\author{Stephen DiAdamo, Janis N\"otzel, Benjamin Zanger, Mehmet Mert Be\c{s}e  \\
    \normalsize{Technische Universit\"at M\"unchen - Lehrstuhl f\"ur Theoretische Informationstechnik}\\
    \normalsize{Theoretical Quantum System Design}\\
    \normalsize{\{stephen.diadamo, janis.noetzel, benjamin.zanger, mehmetmert.bese\}@tum.de}
}

\begin{document}
\maketitle

\begin{abstract}
    As quantum internet technologies develop, the need for simulation software and education for quantum internet rises.  QuNetSim aims to fill this need. QuNetSim is a Python software framework that can be used to simulate quantum networks up to the network layer. The goal of QuNetSim is to make it easier to investigate and test quantum networking protocols over various quantum network configurations and parameters. The framework incorporates many known quantum network protocols so that users can quickly build simulations and beginners can easily learn to implement their own quantum networking protocols.
\end{abstract}

\section{Introduction}\label{sec:intro}

    A quantum internet is a network of devices that are able to transmit quantum information and distribute quantum entanglement amongst themselves. As developments are made towards realising a quantum internet, the first stages of which are likely to be available in the near future \cite{quantum_internet, q_internet_roadmap}, there is a stronger need to be able to efficiently develop and test quantum networking protocols and applications. Recently, there has been much effort into developing quantum simulation software \cite{quantum_libraries}, but much of this effort is directed at quantum computing where fewer efforts have been given to simulation software for quantum networks. A need has therefore arisen for advanced quantum network simulation tools. The initial release of QuNetSim \cite{qunetsim} fills this need by providing a lightweight, easy to use, open-source, network simulator agnostic to any particular quantum network architecture.
    
    As it has been done for the classical internet with, for example, the ns-3 and mininet platforms \cite{ns3, mininet}, work towards a similar open-source simulation platform with many contributors should be developed for quantum networking. Open-source quantum network simulators, as we will discuss in more detail, exist or will soon exist. Presently, we think there is a gap between network simulators that work on very low level and network simulators that are easy to use and can be used in a testing phase. Although QuNetSim does not attempt to reproduce ns-3 or mininet in their entirety for quantum networks, we aim to provide an open-source simulation platform offering a high degree of freedom to attract contributors so that they can add features enabling more simulation possibilities. The goal for QuNetSim is to provide a high-level framework that allows users to quickly develop quantum networking protocols without having to invest time on purely software related tasks, like developing methods of synchronization, writing thread safe qubit simulations, or repeatedly implementing basic protocols that can be used as building blocks for new protocols. As a consequence of meeting these goals, the learning curve needed to begin developing protocols for quantum networks is flattened, since QuNetSim makes it easier to write them. QuNetSim allows users to create examples of quantum networking protocols that are along the lines of how protocols are developed as a first stage for research papers, which helps to develop protocols as well as educate students of quantum networks. In the examples we provide, we see how there is an almost one-to-one correspondence between how protocols are written in research papers as to how simulations are developed in QuNetSim.

    In this section, we give an overview of QuNetSim and details of its features. We then explain the assumptions we have made of future quantum technologies that QuNetSim relies on. We take only a small set of features for granted. We hope to develop QuNetSim iteratively such that the correspondence of the simulations between hardware that is available in reality becomes stronger over time, while at the same time keeping the current flexibility to try new hypothetical devices that do not yet exist. We finish the section by giving a summary of other simulation software and compare their features to QuNetSim.
    
    In the next section we give a detailed overview of the components of QuNetSim and explain how they interact with each other. These components have strong analogies to those in classical networks but have quantum specific features that one should be aware of. We explain the inner workings of the components and how they are implemented in the software so that one can more easily understand how QuNetSim works. A general principle of the software is that everything should run asynchronously. In a realistic setting, hosts are not aware of when an incoming packet can arrive. We mimic this constraint and enforce that hosts and the network sit idle awaiting incoming packets and then perform an action based on the protocol defined in the packet headers. We simulate this via multi-threading and queue processing. We provide deeper insight in the next section.
    
    The following section of the report provides an in depth explanation of how QuNetSim is designed and operates at a software level. We explain how the parts of the software interact with each other as well as how a user should interact with QuNetSim. QuNetSim comes with a toolbox of quantum networking protocols built in, and so we review these capabilities. In the final section, we provide examples using the QuNetSim framework.

    \subsection{Overview of QuNetSim}
        The goal of QuNetSim, as the name suggests, is to simulate quantum-enabled networks. To this end, we aim to allow for the writing and testing of robust protocols for multi-hop quantum transmission with various network parameters and configurations. QuNetSim allows users to create a network configuration of nodes connected via classical or quantum links and then program the behaviour of each node in the network as they want. QuNetSim then provides the methods to synchronize hosts even when they are all performing their actions independently and asynchronously. Further, QuNetSim comes with many built in protocols such as teleportation, EPR generation, GHZ state distribution and more, over arbitrary network topologies, that make it easier to develop more complex protocols, using the basic ones as a toolbox. It also provides an easy way of constructing a complex network topology such that one can design and test routing algorithms for quantum networks.
        
        QuNetSim uses a network layering model inspired by the OSI model. It naturally incorporates control information together with any payload type, but is open to modifications where control information is explicitly transmitted separate from payload. In a realistic quantum internet, it is likely that this exact layering approach will not be used and it could be that new layers are introduced. We do believe though that the basic layering features of the application, transport, and network layers of any quantum network will certainly have the basic separations of an application running, information getting encoded into some form of packet to be put into the network and which will then be routed through the network to the desired destination. We therefore aim at implementing this behaviour and leave the lower layer behaviours to the simulators better suited for simulating the physical realism of the link and physical layers, for example. 
    
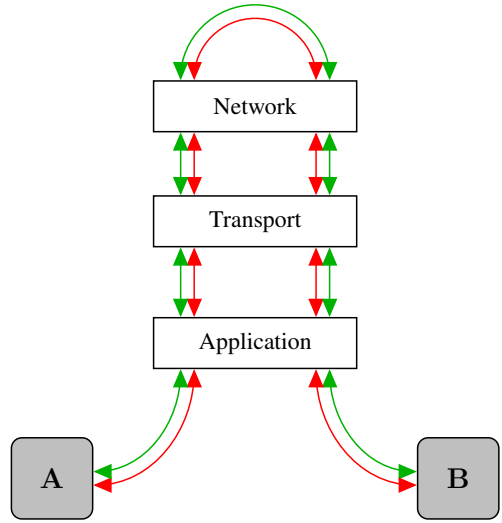
\begin{figure}
        \centering
        \begin{tikzpicture}[scale=0.9, every node/.style={transform shape}]
        \tikzstyle{host}=[draw=black,  rounded corners, line width=0.2mm, fill=lightgray, minimum size=1.2cm]
        \tikzstyle{layer}=[draw=black, line width=0.2mm, minimum width=3cm,  minimum height=0.75cm]
    
        \node (A) [host] at (1, 0) {\large{$\mathbf{A}$}};
        \node (B) [host] at (7, 0) {\large{$\mathbf{B}$}};
        
        \node (app) [layer] at (4, 2) {Application};
        \node (tran) [layer] at (4, 3.8) {Transport};
        \node (net) [layer] at (4, 5.5) {Network};
        
        \path [arrows={triangle 45-triangle 45}, line width=0.2mm, out=0,in=270, red] ([yshift=-0.1cm] A.east) edge ([xshift=-0.9cm]app.south);
        \path [arrows={triangle 45-triangle 45}, line width=0.2mm, out=0,in=270, black!30!green] ([yshift=0.1cm] A.east) edge ([xshift=-1.1cm]app.south);
        
        \path [arrows={triangle 45-triangle 45}, line width=0.2mm,  red] ([xshift=-0.9cm]app.north) edge ([xshift=-0.9cm]tran.south);
        \path [arrows={triangle 45-triangle 45}, line width=0.2mm,  black!30!green] ([xshift=-1.1cm]app.north) edge ([xshift=-1.1cm]tran.south);
        
        \path [arrows={triangle 45-triangle 45}, line width=0.2mm,  red] ([xshift=0.9cm]app.north) edge ([xshift=0.9cm]tran.south);
        \path [arrows={triangle 45-triangle 45}, line width=0.2mm,  black!30!green] ([xshift=1.1cm]app.north) edge ([xshift=1.1cm]tran.south);
        
        \path [arrows={triangle 45-triangle 45}, line width=0.2mm,  red] ([xshift=-0.9cm]tran.north) edge ([xshift=-0.9cm]net.south);
        \path [arrows={triangle 45-triangle 45}, line width=0.2mm,  black!30!green] ([xshift=-1.1cm]tran.north) edge ([xshift=-1.1cm]net.south);
        
        \path [arrows={triangle 45-triangle 45}, line width=0.2mm,  red] ([xshift=0.9cm]tran.north) edge ([xshift=0.9cm]net.south);
        \path [arrows={triangle 45-triangle 45}, line width=0.2mm,  black!30!green] ([xshift=1.1cm]tran.north) edge ([xshift=1.1cm]net.south);
        
        \path [arrows={triangle 45-}, line width=0.2mm, red, out=90, in=180] ([xshift=-0.9cm]net.north) edge (4, 6.8);
        \path [arrows={-triangle 45}, line width=0.2mm, red, out=0, in=90] (4, 6.8) edge ([xshift=0.9cm]net.north);
        
        \path [arrows={triangle 45-}, line width=0.2mm, black!30!green, out=90, in=180] ([xshift=-1.1cm]net.north) edge (4, 7);
        \path [arrows={-triangle 45}, line width=0.2mm, black!30!green, out=0, in=90] (4, 7) edge ([xshift=1.1cm]net.north);
        
        \path [arrows={triangle 45-triangle 45}, line width=0.2mm, black!30!green, out=270, in=180] ([xshift=1.1cm]app.south) edge ([yshift=0.1cm]B.west);
        \path [arrows={triangle 45-triangle 45}, line width=0.2mm, red, out=270, in=180] ([xshift=0.9cm]app.south) edge ([yshift=-0.1cm]B.west);
        \end{tikzpicture}
        \caption{\small{A design depiction of QuNetSim. Here the green line represents a classical channel and the red a quantum channel. QuNetSim attempts to simulate the process of moving both classical and quantum packets through a set of network layers as does the classical Internet. Here Host $\mathbf{A}$ has a virtual connection to  Host $\mathbf{B}$, and so all of their communication is processed one layer at a time. QuNetSim does not explicitly incorporate features of the higher layers like the link-layer or the physical layer.}}
        \label{fig:overview}
        \end{figure}

        In Figure \ref{fig:overview}, we depict the starting point diagram used to design the structure of QuNetSim. At a high level, it resembles a virtual connection between two nodes in a classical network, where ``virtual connection" means that host $\mathbf{A}$ has the perspective that it is directly connected to host $\mathbf{B}$ even though the information sent from $\mathbf{A}$ is routed through the network with potentially many relay hops. In the figure, the two hosts are connected (virtually) by a classical channel, represented by the green lines, and a quantum channel, represented by the red lines. Both modes of communication are processed through the same layering mechanism as the network is able to route both kinds of information but makes decisions based on the payload of the packets. This allows users to use the same logic for sending classical messages as for sending quantum, leaving it to the lower layers to work out the differences.        

        We create a network component for each layer to keep the layers distinct and separated. These components are the host, the transport layer for packet encoding and decoding, and the network itself. The host is responsible for the applications it wants to run. It can run both classical and quantum applications and, therefore, can process both types of information. The transport layer object simply prepares the packets for the network. It completes some of the initial processes that are required in some cases like encoding a qubit for a super-dense message transmission or that handling of generating the two correction bits for quantum teleportation. It also checks if two hosts share an EPR pair before attempting to run a protocol that requires one and runs the EPR creation protocol if not. The network layer contains two parts, namely the quantum and classical part. There are two underlying networks for each part, that is, the network itself is composed of two independent graph structures to represent each network type. In some cases they can be the same, but QuNetSim allows that the connections can be just classical or just quantum as well. The network can route the two types of information using two different routing algorithms if one chooses to configure it as such. The network also handles long distance entanglement distribution using an entanglement swap chain if desired.

        Overall, the current implementation of QuNetSim is not intended to closely simulate the physical properties of a quantum internet. Instead, it aims to simulate the network layer level and above, with future iterations of QuNetSim investigating more deeply into the lower layers. The intended use of QuNetSim is to allow researchers and students to program their quantum network protocols as a first step to developing new applications and routing algorithms without the need for purely software development tasks. As a side-effect, QuNetSim can also be used to introduce those without a strong physics or networking background to quantum networks. We intend the software to be used to create and test new quantum protocols for robustness, but we do not expect the software to be used for bench-marking or testing of physical systems.
        
        In developing QuNetSim, we have made some assumptions of the future of quantum internets. We believe that the features of QuNetSim are at least mostly aligned with the capabilities of what a quantum internet will be able to do. In the next section we cover in depth which assumptions were made and how reasonable they are.
        
    \subsection{Assumptions Made About Future Quantum Networks}\label{sec:assumptions}

        Although much research is directed at building a quantum internet, currently one does not exist nor are the features of a quantum internet yet to be standardized. In order to build a simulation software framework as general as possible while attempting to keep it simple, we therefore make assumptions that we think will be features of near term quantum internet and future generations of them. 
        
        One assumption we make is that quantum information arrives with classical header information. One proposal for a quantum internet is  discussed in the principles of quantum networking document \cite{git_arch_doc}, which states there is no equivalent of a payload carrying packet which carries quantum information with classical control information. Thus control information must be sent separately from the quantum information. To simplify the process of synchronizing the control information being separated from the quantum, we allow for packet header information to be attached to the quantum transmissions. In QuNetSim it is also possible to integrate one's own control information using classical messages and quantum transmissions over two different connections.
        
        Another assumption we make is that quantum nodes will be able to detect that a qubit has arrived without destroying the quantum information. QuNetSim aims to make it easy to synchronize hosts and therefore we need to work with messages being acknowledged. Thus currently, QuNetSim acts as if non-destructive detection was possible. Once probabilities of false positives and false negatives get introduced in future updates, this assumption will be relaxed. We believe that research will bring up methods for almost-perfect nondestructive quantum detection. Experimental results in this direction exist \cite{nondestructive}, but cannot protect the information of any general quantum state. In the initial release we decided to keep the feature, as it simplifies the writing of protocols.
        
        The state of the quantum internet is still very primitive and although we attempted to design QuNetSim in a way that safeguards against future physical implementations, it is still very difficult to estimate how exactly a quantum internet will be designed and implemented and what features it will and will not have. We therefore make no claim that a simulation implemented in QuNetSim is guaranteed to work in a future quantum internet, but we hope that by using it one can more easily envision and experiment with possible quantum networks.
        
    \subsection{Alternative Simulation Software for Quantum Networks}\label{sec:recent_trends}
        There exists many efforts aimed at developing software for simulating quantum systems. A detailed list of quantum software libraries hosted at \cite{quantum_libraries} contains approximately 100 different flavours of quantum simulation software. Most of these are directed at simulating quantum computation and circuitry on various hardware configurations with various levels of realism. With regards to quantum networking, as far as we know at this time, there are three open source (or soon to be open source) quantum network simulators: SimulaQron \cite{simulaqron}, NetSquid \cite{netsquid}, and SQUANCH \cite{squanch}. SimulaQron and SQUANCH are publicly available and open-source. NetSquid, at this time, is not yet publicly accessible, but is expected to be soon. We give a brief summary of the three libraries to compare against the features that QuNetSim provides.
        
        SimulaQron is a simulator that can be used for developing quantum internet software. It simulates several quantum processors that are located at the end nodes of a quantum network and are connected by simulated quantum links. The main purpose of SimulaQron is to simulate the application layer of a network; tasks such as routing are left to the user to implement using their own approach if needed. SimulaQron further offers the ability to run simulations across a distributed system, that is, simulations can be set up to run on multiple computers. What we found slightly lacking in SimulaQron is a way to easily synchronize the parties regarding qubit arrival. A key difference in QuNetSim is that it adds a layer of synchronization. Built into QuNetSim is the approach of acknowledging when information arrives at the receiver. One can more naturally write protocols in a standard way, where one handles the information arriving or not before proceeding. SimulaQron also has hosts which have features like sending qubits, establishing EPR pairs, and sending classical information. To simplify the task of developing protocols on top of existing protocols, we try to include more built-in tasks such as sending teleportation qubits, establishing a GHZ state, and establishing a secret key using QKD.
        
        SQUANCH (Simulator for Quantum Networks and Channels) achieves similar functionality as SimulaQron but allows for customizable physical layer properties and error models. It allows for creating simulations of distributed quantum information processing that can be parallelized for more efficient simulation. It is designed specifically for simulating quantum networks to test ideas in quantum transmission and networking protocols. SQUANCH can be used to simulate many qubits and can allow a user to add their own error models, which we think allows for a more realistic quantum network simulator. SQUANCH also allows one to separate the quantum and classical networks of a complete network and as well as adding length dependent noise to the channel. A key difference between SQUANCH and QuNetSim is that in SQUANCH, as far as we know, a node can run one set of instructions at a time and not more in parallel. This may not be so restrictive, but in multiparty protocols, it may become challenging to develop all of the behaviour in one set of instructions. QuNetSim allows one to develop multiparty protocols one at a time and run them in parallel. Further, synchronization between parties is again potentially an issue with SQUANCH. QuNetSim gives each host an addressable quantum memory such that given an ID, they can fetch a qubit and can manipulate it as desired. In SQUANCH one should initialize their qubits before the start of the simulation whereas with QuNetSim qubits are initialized at run time. We think this adds more flexibility when writing protocols and allows for more natural logic in the code. 
        
        NetSquid is not yet publicly available. Our information about it is from recent publications that made use of the software (see \cite{link_layer}) and from the NetSquid homepage, such as it is. We cannot therefore claim that the description we give is accurate. We have concluded that NetSquid is more like SQUANCH than SimulaQron in that it can simulate the physical properties of quantum devices like link noise and quantum state decoherence with time. We believe that NetSquid is more of a bench-marking tool that mainly simulates the physical and link layer of a quantum network. QuNetSim is designed to more easily develop quantum protocols and test them for robustness over the network, and not for benchmarking against physical properties of the network. This greatly reduces the need to understand quantum networks at a deep level, but does remove the ability to benchmark quantum networking protocols against any hardware specifications.
        
        
        \subsection{Limitations of QuNetSim}
        
        
        QuNetSim provides a high-level framework for developing quantum protocols. There are, however, some limitations in the current implementation. QuNetSim relies on existing qubit simulators (see section \ref{sec:backends}) which in some cases perform well and in some cases do not. This causes QuNetSim to periodically run more slowly than desired and, because we are using full qubit simulators where the alternative is error tracking only, running large scale simulations can consume much of the host computer's resources. We have found that QuNetSim works well for small scale simulations using five to ten hosts that are separated by a small number of hops. QuNetSim tends to reach its limits when many entangled qubits are being generated across the network with many parallel operations. As this is an ongoing project, we will investigate performance improvements as a priority for the next iteration. Moreover, QuNetSim is not aiming at perfect physical realism, and so the physical properties of quantum networks are mostly neglected at this stage.

\section{Design Overview of QuNetSim}\label{sec:design}
    
    The aim of QuNetSim is to allow for the development of simulations that contain enough realism that applications of quantum networks can be developed, tested, and debugged for a proof of principle step. With this in mind, we have designed the software such that we remove the need for the large overhead of setting up new simulations and added built-in features that are repeated across many simulations. Another design aspect we aim for is that a prior deep level of understanding of quantum networks and software development should not be required to use QuNetSim. To allow for as many users as possible to develop their applications, we keep the functionality at a high level such that protocols written with QuNetSim are as easy or easier to understand as the protocols written in scientific papers. We provide the added benefit that such protocols are easily modified and simulated chained together or in parallel under various configurations. QuNetSim follows the classical networking model (i.e. the OSI layer model) thereby easing the transition from classical networking to quantum networking, which helps with education.
    
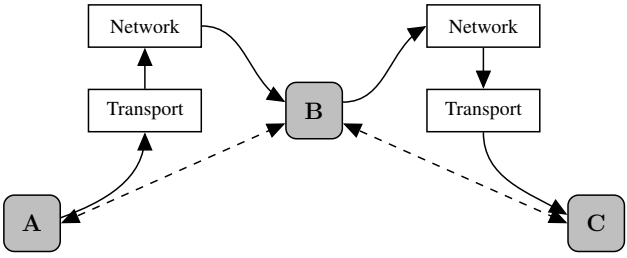
\begin{figure}
    \centering
    \begin{tikzpicture}[scale=0.75, every node/.style={transform shape}]
    \tikzstyle{host}=[draw=black,  rounded corners, line width=0.2mm, fill=lightgray, minimum size=1cm]
    \tikzstyle{layer}=[draw=black, line width=0.2mm, minimum width=2cm,  minimum height=0.75cm]
    
    \node (A) [host] at (0.5, 0.5) {\large{$\mathbf{A}$}};
    \node (B) [host, xshift=5cm, yshift=2cm]  at (0.5, 0.5) {\large{$\mathbf{B}$}};
    \node (C) [host, xshift=10cm, yshift=0cm]  at (0.5, 0.5) {\large{$\mathbf{C}$}};
    \node (transport_A) [layer, xshift=2cm] at (0.5, 2.5) {Transport};
    \node (network_A) [layer, xshift=2cm] at (0.5, 4) {Network};
    \node (network_B) [layer, xshift=8cm] at (0.5, 4) {Network};
    \node (transport_B) [layer, xshift=8cm] at (0.5, 2.5) {Transport};
    
    \draw [arrows={triangle 45-triangle 45}, line width=0.2mm, dashed] (A.east) -- (B);
    \draw [arrows={triangle 45-triangle 45}, line width=0.2mm, dashed] (B) -- (C.west);
    
    \path [arrows={-triangle 45}, line width=0.2mm, out=20,in=270] (A.east)++(0,0.1) edge (transport_A.south);
    \path [arrows={-triangle 45}, line width=0.2mm] (transport_A.north) edge (network_A.south);
    \path [arrows={-triangle 45}, line width=0.2mm, out=0,in=150] (network_A.east) edge ([yshift=1.5mm]B.west);
    
    \path [arrows={-triangle 45}, line width=0.2mm, out=0,in=210] ([yshift=1.5mm]B.east) edge (network_B.west);
    \path [arrows={-triangle 45}, line width=0.2mm] (network_B.south) edge (transport_B.north);
    \path [arrows={-triangle 45}, line width=0.2mm, out=270, in=150] (transport_B.south) edge ([yshift=1.5mm] C.west);
    \end{tikzpicture}
    \caption{\small{An example of a communication process in QuNetSim with three hosts. In this example, there are three hosts, $\mathbf{A}$, $\mathbf{B}$, and $\mathbf{C}$. Hosts $\mathbf{A}$ and $\mathbf{B}$ are connected via a 2-way channel (represented by the dashed line), as are hosts $\mathbf{B}$ and $\mathbf{C}$. When host $\mathbf{A}$ executes an application that transmits information to host $\mathbf{C}$, since there is no direct connection, the information must first be routed though host $\mathbf{B}$. QuNetSim uses a layered approach like the Internet. First, application data is filtered through a transport preparation layer so that the information packet is prepared for the network. From there, the transport layer packet is put into the network. The network also encodes the packet with its own header information and begins to route the packet through the network. The network packet is moved first to host $\mathbf{B}$, and host $\mathbf{B}$ relays the data to host $\mathbf{C}$ to complete the transmission.} }
    \label{fig:network_1}
    \end{figure}

    Figure \ref{fig:network_1} gives an overview of QuNetSim's architecture. Here we see three network nodes, which we call hosts in QuNetSim. Hosts $\mathbf{A}$ and $\mathbf{B}$ are connected via a communication link as are hosts $\mathbf{B}$ and $\mathbf{C}$. As in a classical network, hosts are running such that they sit idle awaiting any incoming packets and then act when packets arrive. Hosts in QuNetSim run applications asynchronously and transfer quantum and classical messages to other hosts in the network. In the figure, host $\mathbf{A}$ runs an application that sends a packet to host $\mathbf{C}$. Host $\mathbf{A}$ has no direct connection to $\mathbf{C}$, and therefore its information must be routed through $\mathbf{B}$ in order to arrive at $\mathbf{C}$. In a layered network architecture, since Host $\mathbf{A}$ is running on the application layer, it should not be concerned with how the information arrives at $\mathbf{C}$, it should just be routed to $\mathbf{C}$ if a route exists in the network. 
    
    The transport layer prepares the information sent from $\mathbf{A}$ for the network by encoding necessary information in a packet header. It should also ensure before putting the packet into the network that the quantum protocols are able to run. For example, it could be that entanglement is needed beforehand. The transport layer ensures this. Once the packet is added to the network the network layer routes it. The path from $\mathbf{A}$ to $\mathbf{C}$ is through $\mathbf{B}$ and so a transport layer packet is encoded in a network packet and then moved through $\mathbf{B}$ to $\mathbf{C}$. When host $\mathbf{B}$ receives a packet from $\mathbf{A}$, since it is not the intended receiver, it relays the network packet onward. Finally, when the network packet arrives at $\mathbf{C}$ the packet is processed through the network layer and then through the transport layer so it can then use the decoded information for her application. This separation of responsibility per layer is a fundamental element of the QuNetSim design.
    
    QuNetSim has three main components, the Host, the Transport, and the Network. Hosts perform analogously to hosts $\mathbf{A}$, $\mathbf{B}$, and $\mathbf{C}$. They run asynchronously and process incoming packets when they arrive. They automatically process incoming packets through the transport layer and receive the payload which is added to a local storage for later processing.
    
    The transport layer processes outgoing and incoming packets, as it does in the Internet. We define a set of protocols such that the packets are encoded and decoded accordingly. A key difference for quantum networks is that some protocols may require that a shared EPR pair is established before the protocol can be performed. We make this a task for the transport layer as well. Once information is encoded into a packet, the transport layer moves the packet to the network. 
    
    The network of QuNetSim behaves much like the network in the Internet with some key differences. In QuNetSim, the network is composed of two internal networks, one for quantum information and one for classical information. When quantum information is sent from a Host, the network routes it through the quantum links in the network as it does for classical information. Another responsibility of the network layer that differs from the classical setting is that the network layer is responsible for establishing an EPR pair between nodes that do not share a direct connection via an entanglement swapping routine. It can trigger a chain of hosts to perform an entanglement swap so that in the end, the sender and the intended receiver will share an EPR pair.
    
    QuNetSim does not currently go above the network layer in terms of simulation of quantum networks. As more features are developed, the code structure allows us to replace pieces that we currently omit, such as link layer functionality. We do, however, allow the user to integrate their qubit channel models, and in subsequent versions will incorporate this into the design such that these things will be easy to change.
    
    \subsection{How QuNetSim Works}
    
    QuNetSim is developed in Python. Naturally, the host in networks processing data sit idle at times, waiting for incoming packets. We make use of Python's threading library to implement this behaviour. Each host is implemented as a queue in a thread. When a host performs an action at the application level, a packet object is generated and put into a packet queue for processing. Transport layer packets contain header information and a payload. The header information contains sender and receiver information, along with information on how to process the payload.
    
    QuNetSim is an event driven simulator where events are triggered by the packets in the network. Host objects have their own packet queue and are monitoring it constantly with some adjustable frequency, checking for new packets in their queue. When a packet is found in the queue, it is processed through the transport layer which is a set of encoding and decoding functions that either prepares packets for the network or unpacks the packets so the host can then access the payload. When the payload is decoded, it is stored in either the classical memory of the host, which is a structured list of Message objects, or in one of the two quantum memories, which is similar to the classical memory but hold qubits that have more parameters. There are two quantum memories as this allows users to more easily distinguish between qubits that contain data and qubits that are entangled. 
    
    The transport layer component of QuNetSim is a set of protocols that run based on the packet object information. It could be that a host sends a classical message encoded via superdense coding using the built in host function for sending superdense encoded messages. For example, internally when a host runs a superdense message transmission protocol, a packet is put onto their own packet queue with packet information stating that the packet is meant to be a superdense encoded message. The transport layer filters this and takes the necessary steps needed such that when the single qubit is transmitted in the network, everything is prepared on the sender and receiver side such that the receiver can then decode the packet correctly. The transport layer of QuNetSim is a processing filter between the host and the network.
    
    Once a packet is encoded at the transport level, it is put into the network and added to the network packet queue. The network acts like a host where it has a packet queue which is being observed for changes. Once a packet is put into the queue, it is analyzed and processed. If the packet is a signal to the network to generate entanglement, there is a setting where the network will conduct an entanglement swap along a chain of nodes done by orchestrating them to establish single hop EPR pairs and then using those pairs to perform a teleportation. If no EPR pair is needed the payload is checked for the type of data it contains. If it is classical data, it is routed over the classical network and if it is quantum data it is routed via the quantum network. A network singleton contains two directed graphs to represent the two networks. By separating the network into two separate graphs it also allows for applying different error models during routing. By default, the routing algorithm is shortest path, but it can be changed via parameter settings. We see an example of how this is done in section \ref{sec:examples}.
    
    The benefit of this code structure and behaviour is that we can produce meaningful log messages that help users debug their protocols. Because each step of a protocol is  executed in the simulation, the log messages show step by step how the protocols are working. This could be especially useful for students who can learn how a protocol works over a complex network step by step.
    
    In summary, QuNetSim implements a layered model of component objects much like the OSI model. The host and network components are implemented using threading and observing queues. The queues are monitored constantly and queue changes trigger and an event. Extensive use of threading allows each task to wait without blocking the main program thread, which simulates the behaviour of sending information and waiting for an acknowledgement, or expecting information to arrive for some period of time from another host. 
    
    QuNetSim supports the use of any particular qubit simulation backend. A point worth noting is that this feature, at times, forces one to slow queue processing time. The right choice of queue processing time depends on the qubit simulation engine. The interaction between QuNetSim and qubit backends is explained in more detail in the next section. 

    \subsection{Network and Qubit Backends}\label{sec:backends}
    
    QuNetSim relies on open-source qubit simulators that we use to simulate the physical qubits in the network. At the current stage, we are using three qubit simulators that each have their own benefits: CQC/SimulaQron \cite{simulaqron}, ProjectQ \cite{projectq}, and EQSN \cite{EQSN}. EQSN is the default backend and it is written by the TQSD group. Users are free to change the backend of QuNetSim to use different qubit simulators and we also explain how new backends can be easily added in our full documentation \cite{qunetsim_docs}.

    The advantage of using ProjectQ is that it is usually very fast. We have found, however, that it quickly slows down when there is a high volume of qubit entanglement. Also, we have observed errors when many measurements are being made on qubits, likely due to threads accessing the engine asynchronously. Generally, these are not major issues and can be avoided by slowing the simulation slightly. Future analysis will find the root of these problems, but careful use of the ProjectQ backend allows it to work well. EQSN and CQC are both working well in terms of threading and processing, but we have observed them to be slower than ProjectQ. They are generally more reliable under many qubit operations and tend not to be affected as quickly when many entangled states are being generated. 
    
    \section{Using QuNetSim}\label{sec:features}
    
    In this section, we introduce the key features for a user of QuNetSim to implement protocols. We list the most commonly used classes which are the qubit, host, and network classes. A full set of documentation is also available at \cite{qunetsim_docs}.
    
    A foundational data structure used in QuNetSim is the qubit. When a qubit is created, it belongs to a specific host and gets assigned a unique ID. It is also possible to choose a custom ID if wanted. A qubit is generated via the following code.
    
\begin{lstlisting}[language=iPython]
 q = Qubit(host, q_id=id)
\end{lstlisting}

    Once a qubit is created by a host, it can be modified and transmitted. To send a qubit to another party, one can send it directly or use a teleportation protocol by using the following two host methods: 
    
    \begin{itemize}
        \item \verb|send_qubit|: Sends a qubit directly
        \item \verb|send_teleport|: Teleports a qubit
    \end{itemize}
    
    Hosts can also establish EPR pairs with another party or a GHZ state with many parties. To do so, the following host methods are in place:

    \begin{itemize}
        \item \verb|send_epr|: Generates and sends an EPR pair to a desired host
        \item \verb|send_ghz|: Generates and sends each piece of a GHZ state to its intended host
    \end{itemize}

    Hosts can send classical messages in three ways: Send an arbitrary string over a classical connection, send binary messages via superdense coding, or classically broadcast messages through the network.
    
    \begin{itemize}
        \item \verb|send_classical|: Send a classical message
        \item \verb|send_superdense|: Send a 2 bit message via super dense coding
        \item \verb|send_broadcast|: Broadcast a message through the network
    \end{itemize}
    
    For synchronization between communicating hosts, it might be beneficial to wait for acknowledgement from the  the receiving host. Waiting for an acknowledgement before proceeding is possible for all sending functions, done by setting a flag in the function called \verb|await_ack|. For example \verb|host.send_superdense('Bob', await_ack=True)|. By setting the flag to false, the host does not wait before executing the actions that follow. Each host has a property to set how long they wait for acknowledgements.
    
    Hosts also can expect an incoming classical message or qubit. When a classical or quantum message arrives, it is stored at the host in its respective memory structure, that is, there is a distinct memory for classical and quantum information. Hosts have the option to fetch the data from their memories so that actions can be performed on it. These functions are:
    
    \begin{itemize}
        \item \verb|get_classical|: Get an ordered list of the received classical messages of a host
        \item \verb|get_next_classical|: Get the next classical message from a host
        \item \verb|get_data_qubit|: Get a data qubit received from a host
        \item \verb|get_epr|: Get an EPR qubit entangled with another host
        \item \verb|get_ghz|: Get an GHZ qubit entangled with some unknown amount of hosts
    \end{itemize}
    
    Much like awaiting acknowledgements, hosts can also wait until something arrives for a fixed amount of time before proceeding. For each ``get" function, there is a parameter \verb|wait=n| where \verb|n| is a floating point number of seconds to wait. For example, \verb|get_epr('Alice', wait=5)| will wait for five seconds for an EPR to arrive from Alice. 
    
    It is expected that quantum memories will be limited to relatively few qubits in the near term. QuNetSim supports limiting the number of qubits stored at a host. The number of EPR qubits and data qubits can be limited separately or a limit for the combined number of qubits can be set. The host methods for setting the limits are:
    \begin{itemize}
        \item \verb|set_epr_memory_limit|: Restricts the number of just stored entanglement qubits 
        \item \verb|set_data_qubit_memory_limit|:  Restricts the number of just stored data qubits 
        \item \verb|memory_limit|:  Restricts the number of stored data qubits and entanglement qubits 
    \end{itemize}
    
    To join hosts together in the network, hosts have methods for adding connections. These methods are 
    \begin{itemize}
        \item \verb|add_connection|: Adds both a classical and quantum connection to a host
        \item \verb|add_c_connection|: Adds a classical connection to a host
        \item \verb|add_q_connection|: Adds quantum connection to a host
    \end{itemize}
    
    Connections can also be removed at run time with the \verb|remove_connection|, \verb|remove_c_connection|, and \verb|remove_q_connection|. One needs only to call the network \verb|network.update_host| to propagate the changes to the network.
    
    The last step is to start the host listening for incoming packets using the \verb|host.start| method. Once started, hosts can be made to run specific protocols, or sets of instructions, using the \verb|host.run_protocol| method which takes a function as a parameter along with the function parameters as we will see examples of in the next section. Running a protocol can be made to block, or a thread is returned to be handled as wanted using the flag \verb|blocking=True|.
    
    Building the network is also part of every simulation. QuNetSim uses the \verb|network| object to abstract the classical and quantum networks. Once the network topology is established between the hosts, hosts are added to the network using the network method \verb|network.add_host|. The network builds a graph structure using the connections of the host to be used for routing. As we will see in the next section, this involves just adding hosts to the network and then calling the \verb|network.start| method. One can also draw the two networks using \verb|network.draw_quantum_network| and \verb|network.draw_classical_network| methods.

    \section{Examples}\label{sec:examples}
    
    We now demonstrate through examples how QuNetSim works and how to write simulations with it. We show that QuNetSim has a very high-level structure with many features in the background that handle many of the complicated tasks such as synchronizing different hosts, establishing entangled states, and can perform complex tasks with just one line of code. This greatly reduced the need for a strong software development background and allows users to quickly build their simulations without worrying about rote programming related tasks.
    
    \subsection{Sending Data Qubits}
\begin{figure}
        \centering
        \begin{tikzpicture}[scale=0.8, every node/.style={transform shape}, router/.pic={ 
            \node () [draw=black, circle, line width=0.2mm, rounded corners, minimum size=0.7cm] at (0, 0) {};
            \draw (0.2, 0.2) -- (0.1, 0.2);
            \draw (0.2, 0.2) -- (0.2, 0.1);
            \draw (-0.2, -0.2) -- (-0.1, -0.2);
            \draw (-0.2, -0.2) -- (-0.2, -0.1);
            \draw (0.2, -0.2) -- (0.1, -0.2);
            \draw (0.2, -0.2) -- (0.2, -0.1);
            \draw (-0.2, 0.2) -- (-0.2, 0.1);
            \draw (-0.2, 0.2) -- (-0.1, 0.2);}]
            \tikzstyle{host}=[draw=black,  rounded corners, line width=0.2mm, fill=lightgray, minimum size=1cm]

            \node (A) [host]  at (0, 6) {A}; 
            \node (B) [host]  at (0, 4) {B}; 
            \node (C) [host]  at (0, 2) {C}; 
            \node (D) [host]  at (0, 0) {D}; 
        
            \path [arrows={triangle 45-triangle 45}, line width=0.2mm, red] ([xshift=-1mm]A.south) edge ([xshift=-1mm]B.north);
            \path [arrows={triangle 45-triangle 45}, line width=0.2mm, red] ([xshift=-1mm]B.south) edge ([xshift=-1mm]C.north);
            \path [arrows={triangle 45-triangle 45}, line width=0.2mm, red] ([xshift=-1mm]C.south) edge ([xshift=-1mm]D.north);
            \path [arrows={triangle 45-triangle 45}, line width=0.2mm, black!30!green] ([xshift=1mm]A.south) edge ([xshift=1mm]B.north);
            \path [arrows={triangle 45-triangle 45}, line width=0.2mm, black!30!green] ([xshift=1mm]B.south) edge ([xshift=1mm]C.north);
            \path [arrows={triangle 45-triangle 45}, line width=0.2mm, black!30!green] ([xshift=1mm]C.south) edge ([xshift=1mm]D.north);

        \end{tikzpicture}
        \caption{ The network depiction for Example A. }
        \label{fig:data_net}
    \end{figure}
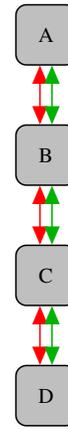

        In this example we demonstrate a simple task of sending qubits that have been encoded with information, or ``data" qubits. We send the qubits from Alice to Dean over the network in Figure \ref{fig:data_net}.
        
\begin{lstlisting}[language=iPython]
 # Network is a singleton 
 network = Network.get_instance()
 # The nodes in the network
 nodes = ["Alice", "Bob", "Eve", "Dean"]
 # Start the network with the nodes defined above
 network.start(nodes)    

 # Define the hosts
 # Note: the host names must match the names above
 host_alice = Host("Alice")
 # Define the host's connections
 host_alice.add_connection("Bob")
 # Start the host
 host_alice.start()

 host_bob = Host("Bob")
 host_bob.add_connections(["Alice", "Eve"])
 host_bob.start()

 host_eve = Host("Eve")
 host_eve.add_connections(["Bob", "Dean"])
 host_eve.start()

 host_dean = Host("Dean")
 host_dean.add_connection("Eve")
 host_dean.start()

 # Add the hosts to the network to build the network 
 # graph
 network.add_hosts([host_alice, host_bob, 
                        host_eve, host_dean])
\end{lstlisting}
    Next, we want to generate the protocols for Alice and Dean to run. Protocols are the functionality of a Host. Protocols are very flexible with the only exception that the protocol function must take the host as the first parameter. Below is sample protocol and code to launch the protocol for a host. In this example, the host, Alice, is sending five data qubits to Dean.

\begin{lstlisting}[language=iPython]
 def sender(host, receiver):
  """
  Sends 5 qubits to host *receiver*.
  Args:
    host (Host): The host object running the protocol   
    receiver (str): The name of the receiver   
  """ 
  for i in range(5):
    # The host creates a qubit
    qubit = Qubit(host)
    # Perform a Hadamard operation on the qubit
    qubit.H()
    # The host sends the qubit to the receiver
    # and awaits an ACK from the receiver that 
    # the qubit arrived for some fixed amount of time.
    ack_arrived = host.send_qubit(receiver, qubit, 
                                            await_ack=True)
    if ack_arrived:
        print('Qubit sent successfully.')
    else:
        print('Qubit did not transmit.')

\end{lstlisting}
    A protocol for receiving qubits must also be written. 
    
\begin{lstlisting}[language=iPython]
 def receiver(host, sender):
  """
  Sends 5 qubits to host *receiver*.
  Args:
    host (Host): The host object running the protocol   
    receiver (str): The name of the sender   
  """ 
  for i in range(5):
    # The host awaits a data qubit for 10 seconds maximum
    qubit = host.get_data_qubit(sender, wait=10)
    # If the qubit arrived, measure it
    if qubit is not None:
        m = qubit.measure()
        print("%s received qubit in state %d" 
                        % (host.host_id, m))
    else:
        print("Qubit did not arrive.")
\end{lstlisting}    
    To run the protocols, we have the following lines of code: 

\begin{lstlisting}[language=iPython]
 # Alice runs the sender protocol and takes     
 host_alice.run_protocol(sender, (host_dean.host_id,))
 host_dean.run_protocol(receiver, (host_alice.host_id,))
\end{lstlisting}    
    
    In summary, with these code snippets we can simulate the transmission of five qubits from Alice to Dean over the network in Figure \ref{fig:data_net}. Of course, this simple example is intended to give an gentle introduction into how QuNetSim works. In the next examples, we develop more complex protocol simulations to see further the simplicity of using QuNetSim to write quantum network simulations.
    
    \subsection{GHZ-based Anonymous Entanglement}
    In this example, we will demonstrate how to simulate an instance of GHZ-based quantum anonymous entanglement \cite{ghz_anon}. The goal of the protocol is to hide the creation of an entangled pair between two parties. This protocol implementation demonstrates the simplicity of translating a protocol from a high-level mathematical syntax into simulation using QuNetSim. The protocol involves establishing GHZ states amongst $n$ parties as well as broadcasting measurement outcomes. These types of tasks would involve a relatively high level of software synchronization in order to program a simulation from scratch. Here we demonstrate that this kind of synchronization logic is kept at a high level.
    
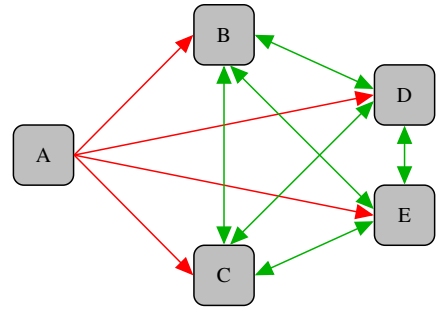
\begin{figure}
        \centering
        \begin{tikzpicture}[scale=0.8, every node/.style={transform shape}, router/.pic={ 
            \node () [draw=black, circle, line width=0.2mm, rounded corners, minimum size=0.7cm] at (0, 0) {};
            \draw (0.2, 0.2) -- (0.1, 0.2);
            \draw (0.2, 0.2) -- (0.2, 0.1);
            \draw (-0.2, -0.2) -- (-0.1, -0.2);
            \draw (-0.2, -0.2) -- (-0.2, -0.1);
            \draw (0.2, -0.2) -- (0.1, -0.2);
            \draw (0.2, -0.2) -- (0.2, -0.1);
            \draw (-0.2, 0.2) -- (-0.2, 0.1);
            \draw (-0.2, 0.2) -- (-0.1, 0.2);}]
            \tikzstyle{host}=[draw=black,  rounded corners, line width=0.2mm, fill=lightgray, minimum size=1cm]

            \node (B) [host]  at (-3, 2) {B};   
            \node (C) [host]  at (-3, -2) {C};   
            \node (D) [host]  at (0, 1) {D};   
            \node (E) [host] at (0, -1) {E};   
            
            \node (A) [host]  at (-6, 0) {A};   
        
            \path [arrows={-triangle 45}, line width=0.2mm, red] (A.east) edge (B.west);
            \path [arrows={-triangle 45}, line width=0.2mm, red] (A.east) edge (C.west);
            \path [arrows={-triangle 45}, line width=0.2mm, red] (A.east) edge (D.west);
            \path [arrows={-triangle 45}, line width=0.2mm, red] (A.east) edge (E.west);
            
            \path [arrows={triangle 45-triangle 45}, line width=0.2mm,  black!30!green] (B.south) edge (C.north);
            \path [arrows={triangle 45-triangle 45}, line width=0.2mm,  black!30!green] ([xshift=1mm] B.south) edge ([yshift=1mm]E.west);
            \path [arrows={triangle 45-triangle 45}, line width=0.2mm,  black!30!green] (B.east) edge ([yshift=1mm]D.west);
            \path [arrows={triangle 45-triangle 45}, line width=0.2mm,  black!30!green] (C.east) edge ([yshift=-1mm]E.west);
            \path [arrows={triangle 45-triangle 45}, line width=0.2mm,  black!30!green] (D.south) edge (E.north);
            \path [arrows={triangle 45-triangle 45}, line width=0.2mm,  black!30!green] ([xshift=1mm] C.north) edge ([yshift=-1mm]D.west);

        \end{tikzpicture}
        \caption{ The network depiction for Example B. }
        \label{fig:ghz_net}
    \end{figure}

    As a first step, as always, we generate a network. Below is the code to generate such a network which is depicted in Figure \ref{fig:ghz_net}.
    
\begin{lstlisting}[language=iPython]
 network = Network.get_instance()
 nodes = ['A', 'B', 'C', 'D', 'E']
 network.start(nodes)

 host_A = Host('A')
 host_A.add_connections(['B', 'C', 'D', 'E'])
 host_A.start()
 host_B = Host('B')
 host_B.add_c_connections(['C', 'D', 'E'])
 host_B.start()
 host_C = Host('C')
 host_C.add_c_connections(['B', 'D', 'E'])
 host_C.start()
 host_D = Host('D')
 host_D.add_c_connections(['B', 'C', 'E'])
 host_D.start()
 host_E = Host('E')
 host_E.add_c_connections(['B', 'C', 'D'])
 host_E.start()
 network.add_hosts([host_A, host_B, host_C, 
                                    host_D, host_E])
\end{lstlisting}  

The next step is to write the behaviour of the GHZ distributor, which in this example is node $A$. QuNetSim provides a function for distributing GHZ states so the function \verb|distribute| simply takes the distributing host as the first parameter and the list of receiving nodes as the second. One notices that the flag \textbf{distribute} has been set to true in the \verb|send_ghz| method call. This tells the sending host that it should not keep part of the GHZ state, rather, it should generate a GHZ state amongst the list given and send it to the parties in the node list, keeping no part of the state for itself.

\begin{lstlisting}[language=iPython]
 def distribute(host, nodes):
    """
    Args:
        host (Host): The host running the protocol
        nodes (list): The list of nodes to distribute the 
                      GHZ to
    """
    # distribute=True => don't keep part of the GHZ
    host.send_ghz(nodes, distribute=True)
\end{lstlisting}

The next type of behaviour we would like to simulate is that of a node in the group that is not attempting to establish an EPR pair. In the anonymous entanglement protocol, the behaviour of such a node is to simply receive a piece of a GHZ state, perform a Hadamard operation on the received qubit, measure it, and broadcast to the remaining participating parties the outcome of a measurement in the computational basis. Below we see how to accomplish this. In the \verb|node| function, the first parameter is, as always, the host that is performing the protocol. The second is the ID of the node distribution the GHZ state, which in this example, is host $A$. The host fetches its GHZ state where, if it is not available at the time, they will wait ten seconds for it, accomplished by setting the \textbf{wait} parameter. If they have not received part of the GHZ state, then the protocol has failed, otherwise they simply perform the Hadamard operation on the received qubit, measures it, and broadcasts the message to the network. QuNetSim includes the task of broadcasting as a built-in function and therefore the task of sending classical messages to the whole network is not only simplified, but the host performing the network does not need to know the entire network to send a broadcast.

\begin{lstlisting}[language=iPython]
 def node(host, distributor):
    """
    Args:
        host (Host): The host running the protocol
        distributor (str): The ID of the distributor 
                           for GHZ states
    """
    q = host.get_ghz(distributor, wait=10)
    if q is None:
        print('failed')
        return
    q.H()
    m = q.measure()
    host.send_broadcast(str(m))
\end{lstlisting}

We implement next the behaviour of the party in the protocol acting as one end of the EPR link that we label the \textit{sender}. Here we take three parameters (other than the host running the protocol), the ID of the distributor, the receiver that is the holder of the other half of the EPR pair, and an agreed upon ID for the EPR pair that will be generated. In QuNetSim, qubits have IDs for easier synchronization between parties. For EPR pairs and GHZ states, qubits share and ID, that is, the collection of qubits would all have the same ID. This is done so that when parties share many EPR pairs, they can easily synchronize their joint operations. The \verb|sender| protocol is the following: first they receive part of a GHZ state, they select a random bit and then broadcast the message so that they appear as just any other node. They then manipulate their part of their part of the GHZ state according to what the random bit was. If the bit was $1$, then a $Z$ gate is applied. The sending party can then add the qubit as an EPR pair shared with the receiver. This EPR pair can then be used as if the sender and receiver established an EPR directly. 

\begin{lstlisting}[language=iPython]
 def sender(host, distributor, receiver, epr_id):
    """
    Args:
        host (Host): The host running the protocol
        distributor (str): The ID of the distributor 
                           for GHZ states
        receiver (str): Who to teleport the qubit to after 
                        EPR is established
        epr_id (str): The ID for the EPR pair established 
                      ahead of time
    """
    q = host.get_ghz(distributor, wait=10)
    b = random.choice(['0', '1'])
    host.send_broadcast(b)
    if b == '1':
        q.Z()

    host.add_epr(receiver, q, q_id=epr_id)
    qubit_to_send = Qubit(host)
    host.send_teleport(r, qubit_to_send)
    host.empty_classical()
\end{lstlisting}

Finally, we establish the behaviour of the \textit{receiver}. The receiver here behaves as follows: First, in order to mask their behaviour they randomly choose a bit and broadcast it to the network. Once complete, they await the remainder of the broadcast messages. In QuNetSim, classical messages are stored as a list in the \verb|classical| field. Since there are 3 other parties, other than the receiver themself, they await the other three messages. Once they arrive, the receiver computes a global parity operation by taking the $XOR$ of all received bits along  with their own random choice. With this, the receiver can apply a controlled $Z$ gate which establishes the EPR pair with the correct sender. They simply add the EPR pair and complete the protocol.

\begin{lstlisting}[language=iPython]
 def receiver(host, distributor, sender, epr_id):
    q = host.get_ghz(distributor, wait=10)
    b = random.choice(['0', '1'])
    host.send_broadcast(b)

    messages = []
    # Await broadcast messages from all parties
    while len(messages) < 3:
        messages = host.classical
        
    parity = int(b)
    for m in messages:
        if m.sender != s:
            parity = parity ^ int(m.content)
    if parity == 1:
        q.Z()

    # Established secret EPR, add it
    host.add_epr(sender, q, q_id=epr_id)
    
    # Await a teleportation from the anonymous sender
    q = host.get_data_qubit(s, wait=10)
    
\end{lstlisting}

The last step of writing a QuNetSim simulation is to run the protocols for each desired host. Below, we let host \textbf{A} act as the GHZ state distributor, \textbf{B} and \textbf{C} are neutral parties running the \verb|node| behaviour, \textbf{D} acts as the sender and \textbf{E} acts as the receiver. The following code initiates the simulation.

\begin{lstlisting}[language=iPython]
 epr_id = '12345'
 host_A.run_protocol(distribute, (['B', 'C', 'D', 'E'],))
 host_B.run_protocol(node, ('A',))
 host_C.run_protocol(node, ('A',))
 host_D.run_protocol(sender, ('A', 'E', epr_id))
 host_E.run_protocol(receiver, ('A', 'D', epr_id))
\end{lstlisting}  
        
    \subsection{Routing with Entanglement}
\begin{figure}
        \centering
        \begin{tikzpicture}[scale=0.8, every node/.style={transform shape}, router/.pic={ 
            \node () [draw=black, circle, line width=0.2mm, rounded corners, minimum size=0.7cm] at (0, 0) {};
            \draw (0.2, 0.2) -- (0.1, 0.2);
            \draw (0.2, 0.2) -- (0.2, 0.1);
            \draw (-0.2, -0.2) -- (-0.1, -0.2);
            \draw (-0.2, -0.2) -- (-0.2, -0.1);
            \draw (0.2, -0.2) -- (0.1, -0.2);
            \draw (0.2, -0.2) -- (0.2, -0.1);
            \draw (-0.2, 0.2) -- (-0.2, 0.1);
            \draw (-0.2, 0.2) -- (-0.1, 0.2);}]
            \tikzstyle{host}=[draw=black,  rounded corners, line width=0.2mm, fill=lightgray, minimum size=1cm]
            
            \pic (A)  at (0, 0) {router};
            \pic (D) at (0, 2) {router};
    
            \node (Alice) [host] at (-3, 1) {A};   
            \node (Bob) [host]  at (3, 1) {B};

            \path [arrows={triangle 45-triangle 45}, line width=0.2mm, out=0,in=180, red] ([yshift=4mm] Alice.east) edge ([yshift=1mm]D.west);
            \path [arrows={triangle 45-triangle 45}, line width=0.2mm, out=0,in=180, black!30!green] ([yshift=2mm] Alice.east) edge ([yshift=-1mm] D.west);
            
            \path [arrows={triangle 45-triangle 45}, line width=0.2mm, out=0,in=180, black!30!green] ([yshift=-2mm] Alice.east) edge ([yshift=1mm]A.west);
            \path [arrows={triangle 45-triangle 45}, line width=0.2mm, out=0,in=180, red] ([yshift=-4mm] Alice.east) edge ([yshift=-1mm]A.west);
            
            \path [arrows={triangle 45-triangle 45}, black!30!green, line width=0.2mm, out=0, in=180] ([yshift=1mm] A.east) edge ([yshift=-0.2cm]Bob.west);
            \path [arrows={triangle 45-triangle 45}, red, line width=0.2mm, out=0, in=180] ([yshift=-1mm] A.east) edge ([yshift=-0.4cm]Bob.west);
            
            \path [arrows={triangle 45-triangle 45}, red, line width=0.2mm, out=0, in=180] ([yshift=1mm] D.east) edge ([yshift=0.4cm]Bob.west);
            \path [arrows={triangle 45-triangle 45}, black!30!green, line width=0.2mm, out=0, in=180] ([yshift=-1mm] D.east) edge ([yshift=0.2cm]Bob.west);
        
        \end{tikzpicture}
        \caption{ The network depiction for Example C. }
        \label{fig:route_net}
    \end{figure}
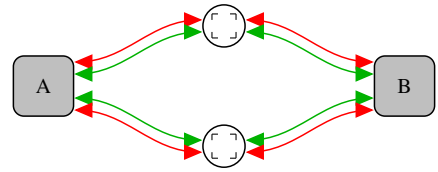

        In this example, we demonstrate how one can use QuNetSim to test a custom routing algorithm. We consider the network shown in Figure \ref{fig:route_net}. For this example, we assume the network is using entanglement resources to transfer classical information via superdense coding from host \textbf{A} to host \textbf{B}. The sending and receiving parties must first establish an EPR pair to send messages via superdense coding. The sender performs a specific set of operations on its half of the EPR pair and then transmits it through the network. When the receiver received the qubit, it performs a specific set of operations such that it recovers two bits of classical information. What is important here is that \textbf{A} and \textbf{B} are separated by one hop. In order to share an EPR pair, an entanglement swap routing has to be made. 
        The strategy for this routing algorithm is to first build a graph of the entanglement shared amongst the hosts in the network. The strategy, since superdense coding consumes entanglement pairs, will then be to route information through the path that contains the most entanglement. In this example, we show how this can be accomplished. 
        As always, we first generate the network topology.
        
\begin{lstlisting}[language=iPython]
 nodes = ['A', 'node_1', 'node_2', 'B']
 network.use_hop_by_hop = False
 network.use_ent_swap = True
 network.set_delay = 0.1
 network.start(nodes)  

 A = Host('A')
 A.add_connections(['node_1', 'node_2'])
 A.start()

 node_1 = Host('node_1')
 node_1.add_connections(['A', 'B'])
 node_1.start()

 node_2 = Host('node_2')
 node_2.add_connections(['A', 'B'])
 node_2.start()

 B = Host('B')
 B.add_connections(['node_1', 'node_2'])
 B.start()

 network.add_hosts([A, node_1, node_2, B])
\end{lstlisting}     

\begin{lstlisting}[language=iPython]
 def generate_entanglement(host):
  """
  Generate entanglement if the host is idle.
  """
  while True:
    # Check if the host is not processing
    if host.is_idle():
      for connection in host.quantum_connections:
         host.send_epr(connection)
           
\end{lstlisting}     

\begin{lstlisting}[language=iPython]
 def routing_algorithm(network_graph, source, target):
  """
  Entanglement based routing function.

  Args:
     network_graph (networkx.DiGraph): The directed graph 
                                       representation of 
                                       the network.
     source (str): The sender ID
     target (str): The receiver ID
  Returns:
     (list): The route ordered by the steps in the route.
  """

  # Generate entanglement network
  entanglement_network = nx.DiGraph()
  nodes = network_graph.nodes()
  # A relatively large number
  inf = 1000000
  for node in nodes:
    host = network.get_host(node)
    for connection in host.quantum_connections:
      num_epr_pairs = len(host.get_epr_pairs(connection))
      if num_epr_pairs == 0:
          entanglement_network.add_edge(host.host_id, 
                                          connection, 
                                          weight=inf)
      else:
          entanglement_network.add_edge(host.host_id, 
                                        connection, 
                                weight=1. / num_epr_pairs)

  try:
    return nx.shortest_path(entanglement_network,
                             source,
                             target,
                             weight='weight')
  except Exception as e:
    print('Error getting route.')
\end{lstlisting}

We can now begin to simulate the network using this configuration. In this simulation, the nodes in the middle are sending entanglement to the other nodes as often as they can, establishing the most EPR pairs while they are free to do so. We start them on this process as so:

\begin{lstlisting}[language=iPython]
 node_1.run_protocol(generate_entanglement)
 node_2.run_protocol(generate_entanglement)
\end{lstlisting}

Now we tell the network to use a different routing algorithm for the quantum information in the network:

\begin{lstlisting}[language=iPython]
 network.quantum_routing_algo = routing_algorithm
\end{lstlisting}

Finally we trigger host \textbf{A} to begin transmitting 100 messages via superdense coding to host \textbf{B}.
\begin{lstlisting}[language=iPython]
 choices = ['00', '11', '10', '01']
 for _ in range(100):
    m = random.choice(choices)
    A.send_superdense('B', m, await_ack=True)
\end{lstlisting}
        
    \subsection{Eavesdropping on Quantum Transmissions}
        \begin{figure}
        \centering
        \begin{tikzpicture}[scale=0.8, every node/.style={transform shape}, router/.pic={ 
            \node () [draw=black, circle, line width=0.2mm, rounded corners, minimum size=0.7cm] at (0, 0) {};
            \draw (0.2, 0.2) -- (0.1, 0.2);
            \draw (0.2, 0.2) -- (0.2, 0.1);
            \draw (-0.2, -0.2) -- (-0.1, -0.2);
            \draw (-0.2, -0.2) -- (-0.2, -0.1);
            \draw (0.2, -0.2) -- (0.1, -0.2);
            \draw (0.2, -0.2) -- (0.2, -0.1);
            \draw (-0.2, 0.2) -- (-0.2, 0.1);
            \draw (-0.2, 0.2) -- (-0.1, 0.2);}]
            \tikzstyle{host}=[draw=black,  rounded corners, line width=0.2mm, fill=lightgray, minimum size=1cm]
            
            \node (Alice) [host] at (0, 0) {A};   
            \node (Eve) [host]  at (3, 0) {E};   
            \node (Bob) [host] at (6, 0) {B};   
            
            \path [arrows={triangle 45-triangle 45}, line width=0.2mm, out=0,in=180, red] ([yshift=1mm]Alice.east) edge ([yshift=1mm]Eve.west);
            \path [arrows={triangle 45-triangle 45}, line width=0.2mm, out=0,in=180, black!30!green] ([yshift=-1mm]Alice.east) edge ([yshift=-1mm]Eve.west);
            
            \path [arrows={triangle 45-triangle 45}, line width=0.2mm, out=0,in=180, red] ([yshift=1mm]Eve.east) edge ([yshift=1mm]Bob.west);
            \path [arrows={triangle 45-triangle 45}, line width=0.2mm, out=0,in=180, black!30!green] ([yshift=-1mm]Eve.east) edge ([yshift=-1mm]Bob.west);
        \end{tikzpicture}
        \caption{ The network depiction for Example D. }
        \label{fig:eve_net}
    \end{figure}
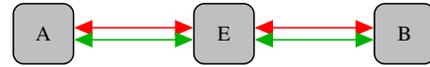
    
    Many quantum networking protocols depend on the detection of qubit manipulation by an eavesdropper. QuNetSim includes the ability to easily add eavesdroppers into the network so that the security of protocols can be tested. In this example, we will see how this is done. We work with the network depicted in Figure \ref{fig:eve_net}. We set up the network as seen above but exclude the code here since it does not differ much from the network in Example A. An attacker in the network is enabled as follows. First, we write the behaviour of the attacker. In this example, the attacker is measuring all of the qubits that pass through. This eavesdropper is also listening to classical messages and appends their own text to every classical message sent. We implement the two functions below. 
    
\begin{lstlisting}[language=iPython]
 def quantum_eve(sender, receiver, qubit):
   """
   Args: 
     sender (str): The sender of the qubit
     receiver (str): The intended receiver of the qubit
     qubit (Qubit): The qubit being transmitted
   """
   qubit.measure(non_destructive=True)

 def classical_eve(sender, receiver, msg):
   """
   Args: 
     sender (str): The sender of the message
     receiver (str): The intended receiver of the message
     msg (Message): The message being transmitted
   """
   msg.content = "I'm listening :)" + msg.content
\end{lstlisting}

Here we define the two functions to take three parameters, which is necessary for QuNetSim. The \verb|sender|, \verb|receiver|, and \verb|qubit| or \verb|msg|. The sender and receiver are the IDs of who sent the qubit and who should receive the qubit respectively. The third parameter is the qubit or message itself. Here one have the flexibility to perform any manipulation on the qubit and message as they choose. In this case  we perform a non-destructive measurement of the qubit which in this case means the qubit is measured, its state collapses, but it stays in the system. Classically this attacker appends text to all classical messages relayed through them.

With this attack, we now configure the eavesdropping host to perform it.
\begin{lstlisting}[language=iPython]
 host_eve.q_relay_sniffing = True
 host_eve.q_relay_sniffing_fn = quantum_eve
 host_eve.c_relay_sniffing = True
 host_eve.c_relay_sniffing_fn = classical_eve
\end{lstlisting}

\textit{Acknowledgement}. This work was funded by the DFG (grant
NO 1129/2-1). 

\clearpage

\small{
}


\begin{thebibliography}{}
    
    \bibitem{qunetsim} DiAdamo, N\"otzel, et al., QuNetSim: GitHub repository, https://github.com/tqsd/QuNetSim
    
    \bibitem{qunetsim_docs}DiAdamo, N\"otzel, et al.,  QuNetSim Documentation: \url{https://tqsd.github.io/QuNetSim/}
    
    \bibitem{quantum_internet} Kimble, H. Jeff. ``The quantum internet." Nature 453.7198 (2008): 1023.
    
    \bibitem{q_internet_roadmap} Wehner, Stephanie, David Elkouss, and Ronald Hanson. ``Quantum internet: A vision for the road ahead." Science 362.6412 (2018): eaam9288. 
    
    \bibitem{qnic} Humble, Travis S., et al. ``Software-defined quantum network switching." Disruptive Technologies in Information Sciences. Vol. 10652. International Society for Optics and Photonics, 2018.
    
    \bibitem{projectq} Steiger, Damian S., Thomas H\"aner, and Matthias Troyer. ``ProjectQ: an open source software framework for quantum computing." Quantum 2 (2018): 49.
    
    \bibitem{EQSN} Zanger, et al., EQSN, Github repository, \url{https://github.com/tqsd/EQSN_python}
    
    \bibitem{nondestructive} Reiserer, Andreas, Stephan Ritter, and Gerhard Rempe. ``Nondestructive detection of an optical photon." Science 342.6164 (2013): 1349-1351.
    
    \bibitem{qtcp} Yu, Nengkun, Ching-Yi Lai, and Li Zhou. ``Protocols for Packet Quantum Network Intercommunication." arXiv preprint arXiv:1903.10685 (2019).
    
    \bibitem{simulaqron} Dahlberg, Axel, and Stephanie Wehner. ``SimulaQron-a simulator for developing quantum internet software." Quantum Science and Technology 4.1 (2018): 015001.
    
    \bibitem{squanch} Bartlett, Ben. ``A distributed simulation framework for quantum networks and channels." arXiv preprint arXiv:1808.07047 (2018).
    
    \bibitem{ns3} Henderson, Thomas R., et al. ``Network simulations with the ns-3 simulator." SIGCOMM demonstration 14.14 (2008): 527.
    
    \bibitem{quantum_libraries} Quantum Open Source Foundation (QOSF), 
    \url{https://qosf.org/project_list/}
    
    \bibitem{netsquid} Coopmans, Dahlberg, et al., NetSquid,  https://netsquid.org/
    
    \bibitem{git_arch_doc} W. Kozlowski, et al. https://datatracker.ietf.org/doc/draft-irtf-qirg-principles/
    
    \bibitem{ghz_anon} Christandl, Matthias, and Stephanie Wehner. ``Quantum anonymous transmissions." International Conference on the Theory and Application of Cryptology and Information Security. Springer, Berlin, Heidelberg, 2005.
    
    \bibitem{link_layer} Dahlberg, Axel, et al. ``A link layer protocol for quantum networks." Proceedings of the ACM Special Interest Group on Data Communication. 2019. 159-173.
    
    \bibitem{mininet} Fontes, Ramon R., et al. ``Mininet-WiFi: Emulating software-defined wireless networks." 2015 11th International Conference on Network and Service Management (CNSM). IEEE, 2015.
    
\end{thebibliography}
\end{document}